\documentclass[conference]{IEEEtran}
\IEEEoverridecommandlockouts

\usepackage{cite}
\usepackage{graphicx}
\usepackage{textcomp}
\usepackage{xcolor}
\usepackage{graphicx,float,wrapfig}
\usepackage{color,soul}
\usepackage{xspace}
\usepackage[utf8]{inputenc}
\usepackage{listings}
\usepackage{xurl}
\usepackage{pifont}
\usepackage{comment}
\usepackage{arydshln}
\usepackage{circledsteps}
\usepackage{fmtcount}
\usepackage{multirow}
\usepackage{threeparttable}
\usepackage{algpseudocode}
\usepackage{amsmath}
\usepackage{fancybox}
\usepackage{tcolorbox}
\usepackage{enumitem}
\usepackage{minted}
\usepackage{caption}
\usepackage{subcaption}
\usepackage[ruled,vlined,linesnumbered]{algorithm2e}
\usepackage{verbdef}
\usepackage{cprotect}
\usepackage{adjustbox}
\usepackage{balance}
\usepackage{tikz}

\def\BibTeX{{\rm B\kern-.05em{\sc i\kern-.025em b}\kern-.08em
		T\kern-.1667em\lower.7ex\hbox{E}\kern-.125emX}}

\lstdefinestyle{mystyle}{
    backgroundcolor=\color{backcolour},   
    commentstyle=\color{codegreen},
    keywordstyle=\color{magenta},
    numberstyle=\tiny\color{codegray},
    stringstyle=\color{codepurple},
    basicstyle=\ttfamily\footnotesize,
    breakatwhitespace=false,         
    breaklines=true,                 
    captionpos=b,                    
    keepspaces=true,                 
    numbers=left,                    
    numbersep=5pt,                  
    showspaces=false,                
    showstringspaces=false,
    showtabs=false,                  
    tabsize=2,
    literate=
     *{0}{{{\color{numb}0}}}{1}
      {1}{{{\color{numb}1}}}{1}
      {2}{{{\color{numb}2}}}{1}
      {3}{{{\color{numb}3}}}{1}
      {4}{{{\color{numb}4}}}{1}
      {5}{{{\color{numb}5}}}{1}
      {6}{{{\color{numb}6}}}{1}
      {7}{{{\color{numb}7}}}{1}
      {8}{{{\color{numb}8}}}{1}
      {9}{{{\color{numb}9}}}{1}
      {:}{{{\color{punct}{:}}}}{1}
      {,}{{{\color{punct}{,}}}}{1}
      {\{}{{{\color{delim}{\{}}}}{1}
      {\}}{{{\color{delim}{\}}}}}{1}
      {[}{{{\color{delim}{[}}}}{1}
      {]}{{{\color{delim}{]}}}}{1}     
}

\def\BibTeX{{\rm B\kern-.05em{\sc i\kern-.025em b}\kern-.08em
    T\kern-.1667em\lower.7ex\hbox{E}\kern-.125emX}}

\newcommand{\sysnamecfi}{{\scshape Watson}\xspace}

\newcommand{\halfcheckmark}{
	\begin{tikzpicture}[scale=0.2]
		\draw[scale=1,fill=black](0,.35) -- (.25,0) -- (1,.7) -- (.25,.15) -- cycle (0.75,0.2) -- (0.77,0.2)  -- (0.6,0.7) -- cycle;
	\end{tikzpicture}
}

\newcommand{\testmark}{
	\begin{tikzpicture}[scale=0.2]
		\draw[scale=1,fill=black](0,.35) -- (.25,0) -- (1,.7) -- (.25,.15) -- cycle;
	\end{tikzpicture}
}

\begin{document}

\title{\sysnamecfi: Leveraging Data Watchpoints for Shadow Stack Protection on Embedded Systems}

 \author{

  \IEEEauthorblockN{Xi Tan}
  \IEEEauthorblockA{%
	\textit{SUNRISE Lab} \\
    \textit{University of Colorado Colorado Springs}\\
	xtan4@uccs.edu
  }
  \and
  \IEEEauthorblockN{Sagar Mohan}
  \IEEEauthorblockA{%
    \textit{CactiLab}\\
    \textit{Northeastern University}\\
   	mohan.sag@northeastern.edu\\
  }
  \and
  \IEEEauthorblockN{Ziming Zhao}
  \IEEEauthorblockA{%
    \textit{CactiLab}\\
    \textit{Northeastern University}\\
   	z.zhao@northeastern.edu\\
  }

}

\maketitle

\begin{abstract}
	
	Embedded and Internet-of-Things (IoT) devices play a critical role in modern life.
	Their software and firmware, often developed in memory unsafe languages like C, are susceptible to memory safety vulnerabilities that can lead to control-flow hijacking attacks.
	Shadow stack is a defense mechanism against control-flow hijacking that targets return addresses.
	However, existing shadow stack solutions for embedded systems have the following limitations. 
	First, they lack system-wide protection, particularly for interrupts and exceptions. 
	Second, they introduce high performance overhead.
	Third, they depend on security extensions like a trusted execution environment, which are not universally available on embedded devices.
	Finally, they rely on hardware features that have inherent constraints, such as limited configurability or resource availability. 
	These restrictions can pose compatibility challenges when integrating additional security mechanisms that require similar hardware support.
	
	To overcome these limitations, we present \sysnamecfi, an efficient and effective shadow stack solution.
	It leverages a standard hardware debug unit named data \underline{wat}chpoints for \underline{s}hadow stack pr\underline{o}tectio\underline{n} on embedded systems.
	To prevent unauthorized access to the shadow stack, \sysnamecfi leverages the address-matching features of the debug unit to enforce the write protection of the shadow stack.
	Only authorized code is allowed to update the shadow stack.
	To prevent forward-edge attack, \sysnamecfi is compatible with compiler options to enforce forward-edge control-flow integrity.
	
	We implemented a prototype of \sysnamecfi on the ARM Cortex-M architecture, and the concept also applies to other platforms.
	The introduced overhead is 7.33\% and 1.81\% on BEEBS and CoreMark-Pro benchmarks, respectively.
	We also evaluate \sysnamecfi on exception handling and two real-world applications, observing negligible performance overhead and a worst-case code size overhead of 2.11\%.
	Furthermore,
	our security evaluation demonstrates that \sysnamecfi effectively prevents attacks that overwrite return addresses and blocks unauthorized modifications to the shadow stack.
	
\end{abstract}

\section{Introduction}
\label{sec:intro}

Memory safety~\cite{szekeres2013sok} remains a critical concern in embedded systems due to the prevalent use of memory-unsafe programming languages like C, which are preferred for their balance between memory access efficiency and portability.
Common vulnerabilities, such as buffer overflows, introduced by these languages, can be exploited to hijack program control flows.
One of the most well-known control-flow hijacking attacks is return-oriented programming (ROP)~\cite{shacham2007geometry, checkoway2010return, davi2010return, bletsch2011jump, bittau2014hacking, weidler2017return}, where an attacker exploits stack-based buffer overflows to overwrite return addresses, redirecting execution to unexpected code sequences.
Given the safety- and security-critical nature of embedded systems, ranging from medical devices to industrial control systems, effectively mitigating such attacks is essential to ensure system reliability and prevent malicious behaviors.

However, microcontrollers that power embedded systems lack the memory protection features found in general-purpose microprocessors, such as the memory management unit (MMU). 
Thus, some general security defense approaches, such as memory virtualization, pose challenges when deployed on embedded systems.
To overcome such hardware limitations, 
researchers have spent over a decade exploring efficient solutions, including inlined instrumentation~\cite{aweke2018usfi, clements2018aces, kim2018securing, mera2022d, shi2022harm, zhou2022opec, khan2023ec, khan2023low, danner2014safer, clements2017EPOXY, tan2024canary, ma2023return, nyman2017cfi, kawada2020tzmcfi, kim2023rio, zhao2024trusted, zhou2020silhouette, du2022kage, walls2019control, richter2024detrap, shao2022faslr} and trace-based monitoring~\cite{tan2023sherloc, wang2024insectacide}.
Inlined instrumentation solutions insert checking code while compiling the source code.
Alternatively, trace-based solutions rely on a hardware trace component to monitor and analyze system execution without access to the source code.
Among all those solutions, shadow stacks are a widely adopted inlined instrumentation solution for protecting backward-edge control flows.
The common implementation is to save the return address in both the main stack and the shadow stack at the function prologue, then retrieve the correct return address from the shadow stack during the function epilogue.
This simple yet robust mechanism can prevent a broad class of backward-edge attacks, significantly strengthening control-flow integrity (CFI).
Obviously, the shadow stack itself must be protected from malicious overwrites. 
Otherwise, attackers could still corrupt the stored return addresses.

Recent research has demonstrated several promising approaches to enforce a \emph{protected} shadow stack on embedded systems.
However, those approaches face both functionality and performance limitations:
i) they do not protect system-wide backward-edge control-flow transfers in the system. 
Specifically, they do not secure interrupt handlers by assuming that the handler functions are trusted.
For example, 
RECFISH~\cite{walls2019control} places the shadow stack at the privileged level, protecting only unprivileged code, leaving privilege exception handler functions unprotected. 
Similarly, 
Silhouette~\cite{zhou2020silhouette} does not safeguard interrupt and exception handlers; and
ii) they introduce high performance overhead~\cite{nyman2017cfi, kawada2020tzmcfi}, which can be up to 513\%.
For example, CFI CaRE~\cite{nyman2017cfi} and TzmCFI~\cite{kawada2020tzmcfi} put the shadow stack in the TrustZone secure state, requiring frequent context switches between the secure and non-secure states to access the shadow stack. 
These context switches cause significant overhead, with increase up to 513\% runtime overhead, 
making such approaches impractical for real-world deployment.

Additionally, existing approaches are constrained by hardware limitations:
i) they~\cite{nyman2017cfi, kawada2020tzmcfi, kim2023rio, zhao2024trusted, tan2023sherloc, wang2024insectacide} rely on security extensions that are not widely available in real-world embedded devices~\cite{tan2024sok}, such as a trusted execution environment (TEE)~\cite{sabt2015trusted}.
As a result, these solutions are impractical for widespread deployment in embedded and IoT devices; 
and
ii) the hardware features these solutions rely on have inherent constraints~\cite{tan2024sok}, limiting compatibility with other security mechanisms.
For example, several projects~\cite{walls2019control, zhou2020silhouette, du2022kage} configure the memory protection unit (MPU)~\cite{Armv8MMPU} to restrict read and write access permissions to the shadow stack.
Since the MPU supports only a limited number of configurable memory regions~\cite{tan2024sok} with a maximum number of eight, reserving regions for a shadow stack could limit the implementation of additional security mechanisms, such as compartmentalization~\cite{aweke2018usfi, clements2018aces, kim2018securing, mera2022d, zhou2022opec, khan2023ec, khan2023low}.

To address those limitations, we propose \sysnamecfi, a compiler-based solution that provides a \textit{protected} shadow stack for embedded systems.
To achieve \textit{compatibility}, \sysnamecfi repurposed a debug unit, namely data watchpoints, which are widely available and featured but rarely used in security mechanisms, as they are intended for debugging during development.
Examples will be data watchpoint and trace unit on ARM Cortex-M~\cite{Armv7MArcRef, Armv8MArcRef} and debug triggers on RISC-V~\cite{RISCVdebugger}.
As a result, \sysnamecfi can work seamlessly with other security mechanisms, including privilege separation and compartmentalization.
Moreover, since \sysnamecfi does not rely on a TEE, it can operate in both the secure and non-secure states, providing control-flow protection for applications in both states.

To ensure \textit{security}, \sysnamecfi enforces write protection on the shadow stack using the address-matching capabilities of data watchpoints. 
Additionally, \sysnamecfi protects the shadow stack pointer by storing it in a memory-mapped register, applying the same write protection mechanism.  
Unauthorized modifications will trigger a system reset or report, while legitimate shadow stack operations do not cause exceptions. 
This is achieved by temporarily granting write access to the shadow stack during function prologues. 

To provide \textit{system-wide} protection of the backward-edge control flows, \sysnamecfi handles interruptions/exceptions and normal executions separately.
This separation guarantees that even malicious exception handlers cannot circumvent the protections in place.
Additionally, \sysnamecfi can be further enhanced by incorporating compilation options, such as \texttt{cfi-icall}, to strengthen the integrity of forward-edge control flows~\cite{llvm_cfi, tice2014enforcing}.

We have implemented \sysnamecfi on the ARM Cortex-M architecture. 
However, the underlying concept of \sysnamecfi can be applied to any architecture that features a hardware debug unit with functionalities similar to data watchpoints.
The implementation enhances the compiler to i) performs static analysis of functions and instruments the function prologue and epilogue to update and retrieve return addresses at runtime,
ii) provides write protection for the shadow stack by utilizing data watchpoints, and 
iii) effectively manages interruptions and exceptions.
We evaluated \sysnamecfi using BEEBS~\cite{pallister2013beebs} and CoreMark-Pro~\cite{coremarkpro2021} benchmarks, applications that trigger exceptions, and two real-world bare-metal systems.
The results demonstrate its \textit{efficiency} with low performance overhead.
Furthermore, we tested its \textit{effectiveness} in defeating attacks that attempt to redirect control flow by rewriting return addresses or modifying return addresses in the shadow stack.

The contributions of this paper are as follows:
\begin{itemize}
	\item We design a \textit{protected} shadow stack for embedded systems, which does not require security extensions or rely on memory protection hardware.
	Thus, \sysnamecfi is compatible with variance existing security mechanisms.
	\item \sysnamecfi ensures system-wide return address integrity for normal functions and exception/interruptions.
	In normal execution, every function call and return updates the main and shadow stacks. 
	For exception handling, \sysnamecfi captures the partial of the exception stack frame at exception entry and checks them against the shadow stack during return.
	\item We implement a prototype of \sysnamecfi and evaluate its \textit{efficiency} and \textit{effectiveness} using various benchmarks. 
	To demonstrate our commitment to open-source, we have made an anonymized version of the code available\footnote{https://anonymous.4open.science/r/Watson-CB18}.
\end{itemize}

\section{Background}
\label{sec:background}

We implement a prototype of \sysnamecfi on ARM Cortex-M architectures, which are designed by ARM for microcontrollers and embedded systems.
In this section, we provide essential background knowledge of the Cortex-M architecture.

\subsection{ARM Cortex-M architectures}
	
\subsubsection{Execution Modes and Privilege Levels}

ARM Cortex-M supports thread and handler execution modes with privileged and unprivileged levels. 
Handler mode always operates in privileged mode, handling exceptions and interrupts.
The thread mode's privilege level can be either privileged or unprivileged, which is determined by the CONTROL register.
The transition from unprivileged to privileged level uses the system supervisor call (\texttt{SVC}).
To switch from the privileged to unprivileged level, one can set CONTROL[0] to 1 through the move-to-system-register (\texttt{MSR}) instruction.

ARM Cortex-M with security extension introduces TrustZone-M, offering physical isolation that splits the system into secure and non-secure states.
The execution modes and privilege levels are further extended into the non-secure and secure states. 
Communication across states is facilitated by new branch and return instructions, maintaining backward compatibility while augmenting security capabilities.

\subsubsection{Exception Model}
\label{subsubsec:exception}

Cortex-M architectures implement hardware stacking and unstacking approaches to make efficient context switches for interrupt/exception handling.
Specifically, when an interrupt occurs, the hardware stacking process automatically pushes the value of related registers, including four registers that store input parameters (\texttt{r0-r3}) passed to the callee, scratch register (\texttt{r12}), current stack pointer (\texttt{r13} or \texttt{SP}), linker register (\texttt{r14} or \texttt{LR}) that stores the return address, and the program counter (\texttt{PC}),
to a stack called an exception stack frame (ESF). 
After that, the \texttt{LR} is updated to a special value (\texttt{EXEC\_RETURN}). 
When the handlers return
that change the \texttt{PC} value with \texttt{EXEC\_RETURN}, the hardware unstacking process is triggered.
The processor first validates the exception states.
Then, the context is restored by popping the ESF back to the relevant registers. 
The \texttt{PC} is updated with the value from the ESF, allowing the program to return to the interrupted execution point.

\subsubsection{Data Watchpoint and Trace Unit}
\label{sec:dwt}

The data watchpoint and trace unit (DWT) features a group of registers called \textit{comparators} that support the matching of instruction/data addresses.
Specifically, DWT can monitor read and/or write accesses to a specified address range.
When the CPU accesses an address that falls within this specified range, it results in a match and generates a watchpoint event. 
This event subsequently triggers the debug monitor (DebugMon) exception, an internal debug mechanism in the processor for handling such debug-related events.
Each comparator within the DWT is composed of a pair of registers: the \texttt{DWT\_FUNCTION} register and the \texttt{DWT\_COMP} register.
These registers are indexed with sequential numbers, allowing for organized configuration and management of the address matching and watchpoint functionalities.
The DWT comparator with ID 0 is commonly used for evaluation and profiling purposes, as it contains the \texttt{DWT\_CYCCNT} register, which can be configured to count CPU cycles or trigger events based on instruction execution. 
In contrast, the other comparators are rarely used.

\begin{table}[!t]
	\centering
	\caption{ARMv7-M DWT registers}
	\footnotesize
	\renewcommand{\arraystretch}{1.2}
	\setlength\tabcolsep{0.8ex}
	\label{tab:dwt-registerv7}
	\begin{threeparttable}
		\begin{tabular}{l|lp{4cm}}
			\hline
			Register & Used fields 	& Description\\ \hline \hline
			\multirow{5}{*}{\texttt{DWT\_FUNCTIONx}}    	
			& DATAVSIZE & Define the data size for value matching \\ \cline{2-3}
			& DATAVMATCH & Control the type of matching  \\ \cline{2-3}
			& \multirow{2}{*}{FUNCTION}  & Define the action on a match and the type of matching  \\ \hline
			\verb|DWT_MASKx|  		
			& MASK   	& Address range matching
			\\ 
			\hline
			\verb|DWT_COMPx|  		
			& COMP   	& Reference value for comparison\\ \hline
			\texttt{DWT\_CYCCNT}    & —      & Counts CPU cycles\\
			\hline
			\hline
		\end{tabular}
	\end{threeparttable}
\end{table}

\begin{table}[!t]
	\centering
	\caption{ARMv8-M DWT registers}
	\footnotesize
	\renewcommand{\arraystretch}{1.2}
	\setlength\tabcolsep{0.8ex}
	\label{tab:dwt-registerv8}
	\begin{threeparttable}
		\begin{tabular}{l|lp{4.4cm}}
			\hline
			Register & Used fields 	& Description\\ \hline \hline
			\multirow{3}{*}{\texttt{DWT\_FUNCTIONx}}    	
			& DATAVSIZE  & Data value size \\ \cline{2-3}
			& ACTION  & Define the action on a match  \\ \cline{2-3}
			& MATCH  & Control the type of match  \\ \hline	
			\verb|DWT_COMPx|  		
			& DADDR   	& Reference value for comparison\\\hline
			\texttt{DWT\_CYCCNT}    & —      & Counts CPU cycles\\
			\hline
		\end{tabular}
	\end{threeparttable}
\end{table}

There are two designs of the DWT unit across Cortex-M architectures. 
Cortex-M architectures prior to Cortex-M23 belong to the ARMv7-M family. 
In these systems, address range matching is configured using a single comparator.
The base address is set via the \texttt{DWT\_COMP} register, and the range size is specified as a power-of-two via the \texttt{DWT\_MASK} register. 
In contrast, the ARMv8-M family uses two comparators.
One to specify the lower bound and another for the upper bound of the address range. 
As a result, systems based on the ARMv8-M family can monitor larger and more flexible memory regions than those based on ARMv7-M. 
Tables~\ref{tab:dwt-registerv7} and~\ref{tab:dwt-registerv8} summarize the key DWT register fields used for address matching in the ARMv7-M and ARMv8-M families, respectively.

\section{System and Threat Model}

\subsection{System Model}
Our prototype targets single-core microcontrollers that feature a debug unit that can be configured for address matching.
Such as the DWT on ARMv7-M and ARMv8-M, and the debug trigger on RISC-V.
The systems running on these microcontrollers are bare-metal systems and do not support multiple threads.
The system may or may not include security mechanisms, such as privilege separation and compartmentalization.  
All system components, including application code, libraries, kernel code, and the hardware abstraction layer (HAL), share the same physical memory address space.
Thus, any memory error in any component can result in a complete compromise of the system.
Same as other CFI solutions, we assume the HAL is trusted due to its critical role in ensuring hardware compatibility.
We also assume access to the system's source code, which is common in inline instrumentation solutions.

\subsection{Threat Model}

We consider an adversary capable of exploiting memory corruption vulnerabilities to perform out-of-bound writes.
Particularly, the adversary can redirect the control flow to either injected or existing code locations (e.g., ROP) by overwriting return addresses. 
The adversary is assumed to have full knowledge of the program's memory contents and layout, including the locations of code, data, and security metadata such as the address of the shadow stack.
We assume the entire system runs either at a single privilege level or separates privileges.
For the formal case, the attacker has the same access rights as the executing system.
In the latter case, the unprivileged attacker may escalate the privilege via buggy interrupt handler functions.
Non-control-data attacks (e.g., attacks that modify data without altering control flow) and physical or hardware-based attacks such as fault injection, power glitching, and electromagnetic tampering are out of scope.
However, \sysnamecfi can defend against attacks that attempt to revise the stored return addresses on the shadow stack.

\section{\sysnamecfi Architecture}
\label{sec:approach}

In this section, we first discuss the security and functional design goals of \sysnamecfi.
We then present an overview of \sysnamecfi's architecture and detailed explanations of its components.

\subsection{Design Goals}

\sysnamecfi has several security and functional design goals.

\begin{itemize}
	\item [G1.] \textit{Return address integrity.} 
	\sysnamecfi should ensure that 
	return instructions always return to a legal address stored by the function prologue.

	\item [G2.] \textit{Interrupt-aware algorithm.}
	\sysnamecfi should prevent potential malicious exception handlers from compromising the system.
	\sysnamecfi should maintain the processor state while entering and returning from an interrupt or exception.
	 
	\item [G3.] \textit{Compatible with existing security mechanisms.}
	\sysnamecfi should be compatible with other existing security mechanisms, allowing a combination of different protections.
	
	\item [G4.] \textit{Low runtime overhead.} 
	\sysnamecfi should be practical and deployable on real-world embedded systems.

\end{itemize}

\subsection{Overview}

\begin{figure}[!t]
	\begin{centering}
		\centering
		\includegraphics[width=0.475\textwidth]{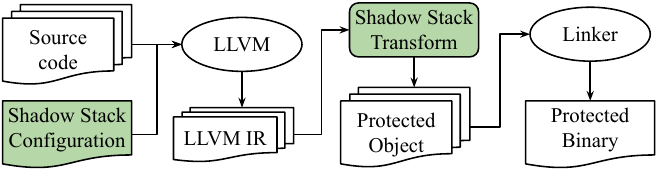}
		\caption{\sysnamecfi Workflow: Components specific to \sysnamecfi are highlighted in green.}
		\label{fig:offline}
	\end{centering}
\end{figure}

As illustrated in Figure~\ref{fig:offline}, \sysnamecfi combines compiler instrumentation with runtime enforcement to safeguard embedded systems.
First, \sysnamecfi transforms the source code of the target system by applying compiler passes to the intermediate representation (IR). 
Then, it links the instrumented objects to create an executable binary.
\sysnamecfi contains three components:
i) \textit{shadow stack transformation} instruments the source code to support the shadow stack, 
ii) \textit{write protection of the shadow stack} configures the hardware debug units to manage the write access to the shadow stack during the runtime,
and iii) \textit{exception protection} handles exception entry and exit separately from normal function calls and returns.

\subsection{Shadow Stack Transformation}

\begin{figure}[!t]
	\noindent\begin{minipage}[t]{.17\textwidth}
		\begin{subfigure}{\linewidth}
			\includegraphics[width=\textwidth]{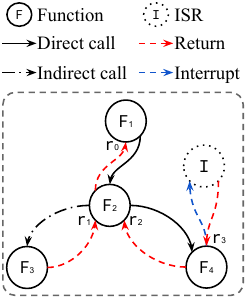}
			\label{fig:fcg}
			\centering
			\subcaptionbox{FCG example}[.9\linewidth]{}
		\end{subfigure}
	\end{minipage}%
	\hfill
	\hfill
	\begin{minipage}[t]{.29\textwidth}
		\begin{subfigure}{\linewidth}
			\includegraphics[width=\textwidth]{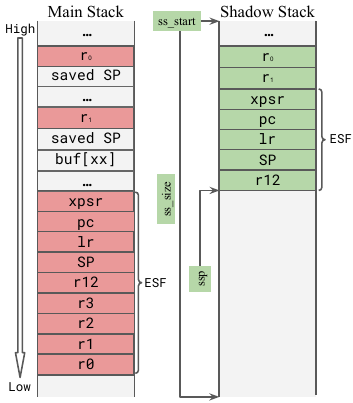}
			\label{fig:ss}
			\centering
			\subcaptionbox{Runtime enforcement}[.9\linewidth]{}   
		\end{subfigure}    
	\end{minipage}
	\caption{\sysnamecfi Overview. 
		(a) illustrates the function call graph of the target system. 
		The control flow transfers include normal function calls and returns, as well as exception entry and exit. 
		(b) illustrates the compact shadow stack.}
\label{fig:overview}
\end{figure}

The \textit{shadow stack transformation} component instruments the prologue of each function to save the return address of function calls to a shadow stack, and instruments the function's epilogue to retrieve the stored address when executing returns, which meets the first design goal (\textbf{G1}).
To make \sysnamecfi compatible with small memory sizes of embedded systems, as Figure~\ref{fig:overview}.(b) shows, 
we use the compact shadow stack, which condenses the size of the shadow stack to store only the return addresses instead of duplicating the whole program main stack used by the parallel shadow stack.

A challenge of using the compact shadow stack is that it requires maintaining a shadow stack pointer (\texttt{ssp}) to access the last entry on the shadow stack, which can be expensive~\cite{burow2019sok}.
One method is to reserve a general-purpose register (GPR).
However, the reserved GPR may be used in other uninstrumented library codes, such as the HAL, resulting in unexpected consequences.
Moreover, on resource-constrained platforms with limited GPRs (e.g., 13 on ARM Cortex-M), reserving one for the entire runtime can lead to increased register pressure and potentially introduce performance overhead.
Thus, the \texttt{ssp} should be an easily accessible register that does not harm the normal execution of the system. 

Consequently, we observed that debug unit registers, while memory-mapped, are seldom utilized in production firmware. 
We can redefine a barely used memory-mapped register as the \texttt{ssp},
such as the registers used to configure the DWT, the instrumentation trace macrocell (ITM), the flash patch and breakpoint unit (FPB), and the trace port interface unit (TPIU).  
However, it raises another issue regarding the protection of the \texttt{ssp} from malicious code access.
To address this, we apply the compiler's forward-edge CFI feature, which restricts indirect branches to valid, statically verified targets.

\subsection{Write Protection of Shadow Stack}
\label{subsec:write}

After transformation, the system uses a shadow stack.
However, the shadow stack is not protected.
As detailed in \S\ref{sec:dwt}, 
we observe that the debug comparators support monitoring of specified address ranges for certain types of access, e.g., read or write, thereby triggering an exception. 
\sysnamecfi utilizes this feature to oversee all write operations to the shadow stack.
Unlike previous work, \sysnamecfi does not require memory protection hardware that limits its compatibility with existing security mechanisms, which meet the third design goal (\textbf{G3}). 

\begin{figure*}[!ht]
	\begin{minipage}[t]{.495\textwidth}
		\begin{subfigure}{\linewidth}
			{\renewcommand\fcolorbox[4][]{\textcolor{gray}{\strut#4}}
				\begin{minted}[xleftmargin=5.8pt,numbersep=1pt, tabsize=2, frame=single, framesep=1mm, breaklines, highlightlines={2-10, 12-18}, linenos=true, escapeinside=||, fontsize=\footnotesize]{tasm}
; assume no free registers
|\xglobal\colorlet{FancyVerbHighlightColor}{green!10}|push   {R4, R10, R12}	; reserve three registers
movw   R12, #4128		  ; lower half of comparator register
movt   R12, #57344	   ; upper half of comparator register
; reconfigure COMPx
movw   R10, #xxxx	    ; lower half of ssp
movt   R10, #xxxx      ; upper half of ssp
; update ss and ssp
str    R4, [R12]       ; store updated ssp back
pop   {R4, R10, R12}	 ; restore reserved reg.
				\end{minted}
			}
			\centering
			\subcaptionbox{Naive instrumented code to access comparator registers and \texttt{ssp}, requiring three free GPRs and seven instruction.}[.9\linewidth]{} 
			\label{lst:acces_naive}  
		\end{subfigure}    
	\end{minipage}
	\begin{minipage}[t]{.495\textwidth}
		\begin{subfigure}{\linewidth}
			{\renewcommand\fcolorbox[4][]{\textcolor{gray}{\strut#4}}
				\begin{minted}[xleftmargin=5.8pt,numbersep=1pt, tabsize=2, frame=single, framesep=1mm, breaklines, highlightlines={2-6, 8-10, 12-13, 16-18}, linenos=true, escapeinside=||, fontsize=\footnotesize]{tasm}
; assume no free registers
|\xglobal\colorlet{FancyVerbHighlightColor}{green!10}|push   {R4, R12}	     ; reserve two registers
movw   R12, #4128		  ; lower half of base addr
movt   R12, #57344	   ; upper half of base addr
; reconfigure COMPx
ldr.w  R4, [R12, #16]  ; get address of ssp
					
; update ss and ssp 
str    R4, [R12, #16]  ; store updated ssp back
pop   {R4, R10}	      ; restore reserved reg.
				\end{minted}
			}
			\centering
			\subcaptionbox{Optimal instrumented code, requiring two free GPRs and six instructions.}[.95\linewidth]{} 
			\label{lst:access_opt}  
		\end{subfigure}    
	\end{minipage}
	\caption{ Code instrumentation for accessing comparator registers and \texttt{ssp}. Instrumented prologue in light green.}
	\label{lst:rt_naive}
\end{figure*}

\begin{figure*}[!ht]
	\begin{minipage}[t]{.46\textwidth}
		\begin{subfigure}{\linewidth}
			{\renewcommand\fcolorbox[4][]{\textcolor{gray}{\strut#4}}
				\begin{minted}[xleftmargin=5.8pt,numbersep=2pt, tabsize=2, frame=single, framesep=1mm, breaklines, highlightlines={2-12}, linenos=true, escapeinside=||, fontsize=\footnotesize]{tasm}
|\xglobal\colorlet{FancyVerbHighlightColor}{green!10}|; assume we have one free register: R12.
push   {R4}           ; reserve register
; load base address to R12
mov.w	R4, #0			   ; scratch register
str.w	R4, [R12, #8]	; disable write protection
ldr.w	R4, [R12, #16] ; load address of ssp
str.w	LR, [R4]			 ; store LR into ss
addw   R4, R4, #4		 ; increment ssp
str.w	R4, [R12, #16] ; store updated ssp
mov.w	R4, #6			   ; scratch register
str.w	R4, [R12, #8]  ; enable write protection
pop    {r4}				   ; restore reserved register
push	 {R6, R7, LR}   ; original instruction
				\end{minted}
			}
			\centering
			\subcaptionbox{Instrumented code for function prologue, in light green. The original instruction is \texttt{push \{R6, R7, LR\}.}}[.95\linewidth]{} 
			\label{fig:ins1_pro}  
		\end{subfigure}    
	\end{minipage}
	\hfill
	\hfill
	\begin{minipage}[t]{.49\textwidth}
		\begin{subfigure}{\linewidth}
			{\renewcommand\fcolorbox[4][]{\textcolor{gray}{\strut#4}}
				\begin{minted}[xleftmargin=5.8pt,numbersep=2pt, tabsize=2, frame=single, framesep=1mm, breaklines, highlightlines={3-13}, linenos=true, escapeinside=||, fontsize=\footnotesize]{tasm}
|\xglobal\colorlet{FancyVerbHighlightColor}{yellow!20}|; assume we have one free register: R12.
; the original instruction pop {R6, R7, PC} is converted into two
|\xglobal\colorlet{FancyVerbHighlightColor}{yellow!20}|pop 	 {R6, R7}   ; restore original saved regs
|\xglobal\colorlet{FancyVerbHighlightColor}{blue!10}|add 	 SP, #4 	  ; balance the main stack (LR)
push   {R4} 		  ; reserve reg			    
; load address of ssp to R12
ldr.w  R4, [R12]	; load ssp value
subw   R4, R4, #4 ; decrement the ssp
ldr.w	LR, [R4]	 ; load return addr				
str.w	R4, [R12]	; store updated ssp
pop 	 {R4}  		 ; restore reserved reg
|\xglobal\colorlet{FancyVerbHighlightColor}{yellow!20}|bx     LR	 		  ; return
				\end{minted}
			}
			\centering
			\subcaptionbox{Instrumented code for function epilogue in light blue. Return case: \texttt{pop \{..., pc\}}. Converted instructions in light yellow.}[.95\linewidth]{} 
			\label{fig:ins1_epi}  
		\end{subfigure}    
	\end{minipage}
	\caption{\sysnamecfi Code instrumentation for function prologue and epilogue.}
	\label{fig:ins1}
\end{figure*}

\subsubsection{Write Protection Initialization} 
The initialization stage sets the address range and access permissions for the shadow stack.
First, \sysnamecfi configures one comparator to store the starting address of the shadow stack, referred to as \texttt{ss\_start}. 
Second, \sysnamecfi sets the address-matching range, referred to \texttt{ss\_size}, by either defining a mask register associated with \texttt{ss\_start} (for ARMv7-M) or configuring an additional comparator to store the shadow stack's ending address (for ARMv8-M).
This range is predefined and remains static throughout the system's operation. 
Third, to activate the write protection, \sysnamecfi i) enables write triggering event for the comparators by configuring associated function registers and ii) activates the DebugMon exception mechanisms by setting the appropriate control bit in the debug exception and monitor control register (DEMCR).
\sysnamecfi modifies the DebugMon exception handler to respond to exceptions triggered by unauthorized write access to the shadow stack, which signals that an attacker attempts to alter the shadow stack.
In response, \sysnamecfi initiates a system reset or sends a report, ensuring the shadow stack's isolation.

However, as the DEMCR is located in the system control block (SCB) region, which is also memory-mapped, attackers are able to access it to alter the active status. 
To avoid this, \sysnamecfi locks the configuration of DEMCR by applying the same write protection of the shadow stack.

\subsubsection{Runtime Enforcement}
\label{subsubsec:rt}
The shadow stack's write protection effectively prevents unintended modifications but also restricts legitimate updates after initialization. 
To accommodate necessary writes, \sysnamecfi temporarily disables write protection during function prologues. 
Specifically, it reconfigures the relevant control register to allow writes, then accesses the current shadow stack entry using the pointer register \texttt{ssp}, which maps to any available memory-mapped register. 
Then, \sysnamecfi writes the return address to the shadow stack and updates \texttt{ssp} to point to the next available slot. 
Before the function body begins execution, write protection is re-enabled.
Upon function return, read access to the shadow stack remains permitted, as the hardware debug unit is configured to restrict only write operations. 
Consequently, the instrumented code in the function's epilogue safely retrieves the return address from the shadow stack via \texttt{ssp}. 
Afterward, \sysnamecfi deallocates the used entry by updating \texttt{ssp}.

These operations require \sysnamecfi to access debug comparator registers and update the \texttt{ssp}, all of which are 32-bit memory-mapped registers. 
Because the instruction set on ARM Cortex-M is also 32-bit.
It cannot encode a full 32-bit immediate value in a single instruction.
Instead, it requires two load instructions (e.g., \texttt{MOVW} and \texttt{MOVT}). 
As a result, Figure~\ref{lst:rt_naive}.(a) shows that accessing these registers, without even modifying their contents, requires three GPRs and incurs a total of seven instructions. 
If these GPRs are not readily available, \sysnamecfi must explicitly reserve and release them for this purpose (Lines 2 and 10). 
While this overhead can be negligible for a single function, it will significantly increase the performance overhead and overall code size when applied across many functions.

To improve performance and satisfy the fourth design goal (\textbf{G4}), we further analyze possible optimization strategies. 
We observe that debug comparator registers are sequentially memory-mapped, with up to four available groups. 
However, only one or two comparators are required for enforcing shadow stack write protection. 
Based on this observation, as shown in Figure~\ref{lst:rt_naive}.(b), \sysnamecfi repurposes an additional comparator register as the \texttt{ssp} and uses a single GPR to hold the base address of the comparator group. 
From this base, \texttt{ssp} is accessed via an offset, thereby reducing the need for one GPR and eliminating redundant move instructions.

Figure~\ref{fig:ins1} presents the full design of the instrumentation.
In the function prologue (Figure~\ref{fig:ins1}.(a)), before executing the original instruction at Line 13, \sysnamecfi temporarily enables write access to the shadow stack (Lines 4-5 and 10-11), stores the return address onto the shadow stack (Lines 6-7), and then increments \texttt{ssp} by 4 bytes to point to the next available entry (Lines 8-9).
Note that while the shadow stack exhibits stack-like behavior, it is implemented as an array and does not follow the traditional address layout of a standard call stack.
In the function epilogue (Figure~\ref{fig:ins1}.(b)), \sysnamecfi simply retrieves \texttt{ssp} to load the return address (Lines 7-9) and updates \texttt{ssp} after use (Line 10).

However, this design may raise concerns because the function prologue embeds instructions that disable the shadow stack's write protection. 
If an attacker exploits a forward-edge vulnerability to launch a jump-oriented programming attack, the attacker may still be able to deactivate \sysnamecfi's protection. 
To mitigate this risk, we enable mature compiler-based mechanisms for enforcing forward-edge control-flow integrity~\cite{tice2014enforcing}.

\begin{listing}[!t]
	{\renewcommand\fcolorbox[4][]{\textcolor{gray}{\strut#4}}
		\begin{minted}[xleftmargin=5.8pt,numbersep=1pt, tabsize=2, frame=single, framesep=1mm, breaklines, highlightlines={2-21, 22-39}, linenos=true, escapeinside=||, fontsize=\footnotesize]{tasm}
; prologue instrumentation - R4 and R5 are free
|\xglobal\colorlet{FancyVerbHighlightColor}{green!10}|; IP stores the addr of base, R4 stores ssp
; disable write protection
ldr.w R5, [SP, #48] ; xPSR values from ESF
str.w R5, [R4]      ; store xPSR to ss
addw R4, R4, #4    
ldr.w R5, [SP, #40] ; RET values from ESP
str.w R5, [R4]      ; store RET to ss
addw R4, R4, #4    
ldr.w R5, [SP, #36] ; LR from ESP
str.w R5, [R4]      ; store LR to ss
addw R4, R4, #4    
ldr.w R5, [SP, #32] ; R12 values from ESP
str.w R5, [R4]      ; store R12 to ss
addw R4, R4, #4    
str.w LR, [R4]      ; current return address
addw R4, R4, #4    
str.w R4, [IP, #16] ; store updated ssp
; enable write protection
; body
; epilogue instrumentation
|\xglobal\colorlet{FancyVerbHighlightColor}{yellow!20}|pop {R7}
|\xglobal\colorlet{FancyVerbHighlightColor}{blue!10}|add sp, #4          ; balance stack
; IP stores the addr of base, R4 stores ssp
mov.w LR, R5        ; restore return address
subw R4, R4, #4     
ldr.w R5, [R4]      
str.w R5, [SP, #28] ; restore R12 to ESP
subw R4, R4, #4     
ldr.w R5, [R4]     
str.w R5, [SP, #32] ; restore LR to ESP
subw R4, R4, #4     
ldr.w R5, [R4]
str.w R5, [SP, #36] ; restore RET to ESP
subw R4, R4, #4     
ldr.w R5, [R4]
str.w R5, [SP, #40] ; restore xPSR
str.w R4, [IP]      ; store updated ssp
|\xglobal\colorlet{FancyVerbHighlightColor}{yellow!20}|bx LR
		\end{minted}
	} 
	\cprotect\caption{Exception handling includes storing and restoring the partial exception stack frame to ensure the integrity of exception control flows.}
	\label{code:exception}
\end{listing}

\subsection{Exception Handling}
\label{subsec:except}

When an exception occurs, control flow is transferred from the normal code to the exception handler function. 
If the exception handler contains vulnerabilities, attackers can exploit them to manipulate the exception stack frame and redirect the return target, as they do with standard function returns. 
If the attacker is operating at an unprivileged level, this could lead to privilege escalation.
To this end, \sysnamecfi is designed to be interrupt-aware to meet the second design goal (\textbf{G2}). 
\sysnamecfi must be compatible with the hardware's native stacking and unstacking mechanisms, as discussed in \S\ref{subsubsec:exception}, 
to minimize performance overhead.

Specifically, when triggering an exception, before executing the exception service routine, \sysnamecfi stores a partial copy of the exception stack frame (ESF) to the shadow stack, omitting parameter-passing registers (e.g., \texttt{R0-R3} on Cortex-M), which are caller-saved and thus not needed for integrity protection. 
This step is illustrated in Figure~\ref{fig:overview}(b) and Listing~\ref{code:exception}, Lines 2–14. 
After saving the ESF, \sysnamecfi stores the current function's return address to the shadow stack and updates \texttt{ssp} (Lines 16–18). 
Before the exception handler returns, the epilogue executes instrumented code to restore the return address and copies the saved ESF from the shadow stack back to the main stack.
By doing so, \sysnamecfi ensures the integrity of both the return address and the ESF while still leveraging hardware-assisted unstacking for efficiency.

Although shadow stack write protection is initialized immediately after the system boots, exceptions may occur due to the system's specific implementation. 
For example, the \texttt{SysTick} exception might be triggered before the memory-mapping registers are initialized. 
At this point, attempting to update the shadow stack, which requires accessing memory-mapped comparators, could lead to unexpected errors.
We consider these exceptions trusted, as they are system-generated and occur before any user input that could cause control-flow violations. 
To address this, \sysnamecfi includes an additional check to ensure the instrumented code accesses the shadow stack only after initialization. 
This is achieved by verifying a specific control bit in the DEMCR or checking the relevant function register of the comparator.

For multi-tasking systems, \sysnamecfi must support context switching. 
Similar to prior work such as Kage~\cite{du2022kage} and TzmCFI~\cite{kawada2020tzmcfi}, \sysnamecfi can modify the \texttt{PendSV} handler, which is responsible for task context switching, to manage shadow stacks on a per-task basis. 
Specifically, each task is assigned a dedicated and fixed-size shadow stack, and \sysnamecfi saves and restores shadow stack state during context switches. 
These per-task shadow stacks are individually protected using hardware comparators, and \sysnamecfi switches comparator configurations accordingly to ensure the integrity of each task's control-flow state.

\section{Implementation and Evaluation}
\label{sec:eva}

We implemented our design by utilizing the LLVM~\cite{llvm_org} compiler and developing a MachineFunction pass to achieve the functionality of \sysnamecfi.
The instrumentation happens after LLVM's instruction selection and register allocation to prevent any unwarranted modifications.

\subsection{Implementation}

We implement our prototype on the Cortex-M4 architecture, though the idea can also be extended to other architectures that support similar functionality as DWT comparators.
The newly implemented LLVM 9.0.1 passes comprise 876 lines of C++ code.
We added code to the Reset and DebugMon handler to support the write protection, which includes initialization and runtime enforcement phases with a total of 56 lines of C code.
Our prototype runs on the STM32F469 Discovery board~\cite{stm32f469}, which is based on the ARMv7-M architecture and includes four DWT comparators.
Each comparator is capable of monitoring an address range up to 32 KB.
Collectively, these four DWT comparators can oversee an address range matching up to 128 KB.
Implementing a parallel shadow stack poses a challenge on this board as it mirrors the main stack in both structure and size, which is of size 2MB.
On the other hand, for the ARMv8-M architecture, we can deploy either the parallel shadow stack or the compact shadow stack.

\emph{Shadow Stack Implementation.}
Table~\ref{tab:dwt-ss} illustrates how we configure the DWT comparators to enforce write protection on different memory regions. 
We implement the compact shadow stack approach using the comparator with ID 0. 
Specifically, \texttt{DWT\_COMP0} holds the base address of the shadow stack, and \texttt{DWT\_MASK0} is set to \texttt{0x15} to define the protected region size.
For comparators configured to monitor write access, we set the associated function registers to \texttt{0x6}, which activates a triggering event when a one-byte write access matches the specified address. 
Although comparator 0 is also used for CPU cycle counting via the \texttt{DWT\_CYCCNT} register, our configuration does not interfere with its normal functionality. 

\begin{table}[ht!]
	\centering
	\caption{Configuration of ARMv7-M DWT registers.}
	\footnotesize
	\renewcommand{\arraystretch}{1.1}
	\setlength\tabcolsep{1.8ex}
	\label{tab:dwt-ss}
	\begin{threeparttable}
		\begin{tabular}{l|llp{2.6cm}}
			\hline
			ID &Register & Value 	& Description\\ \hline \hline
			\multirow{2}{*}{0}&\texttt{DWT\_COMP0} & 0xE00000 & Start address of the ss \\
			&\texttt{DWT\_MASK0} & 0x15 & Size of the ss: $2^{15}$ B = 32 KB \\ \hline
			\multirow{2}{*}{1}&\texttt{DWT\_COMP1} & 0xE00000 & Address of the ssp\\
			&\texttt{DWT\_MASK1} & - & \\ \hline
			\multirow{2}{*}{2}&\texttt{DWT\_COMP2} & 0xE000EDFC & Address of the DEMCR \\
			&\texttt{DWT\_MASK2} & 0x01	& Size of lower 4 bits of DEMCR\\ 
			\hline
			\multirow{2}{*}{3}&\texttt{DWT\_COMP3} & -- & -- \\
			&\texttt{DWT\_MASK3} & -- &  -- \\ \hline
			\multirow{3}{*}{-}&\multirow{3}{*}{\texttt{DWT\_FUNCTIONx}}  &  \multirow{3}{*}{0x6} & Triggering DebugMon exception on matching a one-byte write access\\ 
			\hline
		\end{tabular}
	\end{threeparttable}
\end{table}

By default, the size of a shadow stack is \texttt{32~KB}, as limited by a single comparator's monitoring range. 
To support a larger shadow stack, additional comparators can be configured. 
For instance, comparator ID 3 is reserved and may be used to extend the monitored address range for shadow stack protection. 
Furthermore, the comparator currently used for storing the shadow stack pointer can be reclaimed if an alternative memory-mapped register is adjacent to the DWT registers region.

\emph{Forward-edge CFI Enforcement.}
\sysnamecfi is also compatible with forward-edge CFI mechanisms. 
\sysnamecfi leverages LLVM's compilation option \texttt{-fsanitize=cfi-icall} to provide bounds-checked jump tables and type-based CFI~\cite{abadi2005ccs, abadi2009control}, which restricts indirect branches and calls~\cite{tice2014enforcing}. 
Since determining precise indirect targets remains an open challenge, we do not claim originality in using LLVM for this purpose.

\subsection{Performance Evaluation}

\subsubsection{Experiment Setup}

We use various benchmarks to test \sysnamecfi.
BEEBS~\cite{pallister2013beebs} and CoreMark-Pro~\cite{coremarkpro2021} are used to evaluate \sysnamecfi's performance and code size overhead. 
BEEBS is a lightweight benchmark suite designed to evaluate performance on embedded systems. 
It includes a variety of workloads, such as cryptography, sorting, and matrix operations.
Since some BEEBS programs are too small and would be overly optimized during compilation~\cite{zhou2020silhouette}, we chose 22 programs with more complex execution times.
CoreMark-Pro is a comprehensive benchmark suite that features compute- and memory-intensive workloads, reflecting real-world applications like data compression.

However, BEES and CoreMark-Pro do not include benchmarks that enable interrupts or exceptions. 
Therefore, we developed a custom test case that triggers a \textit{Usage Fault} by attempting to execute an invalid instruction.

We further evaluate \sysnamecfi using two real-world applications. 
Pinlock is a program that toggles a physical LED based on the user's input PIN.  
FatFS\_RAM is an application by STM32Electronics~\cite{stm32} that demonstrates the use of the STM32 firmware to create a FatFS library module. 
FatFS\_RAM is an application that creates a FatFS library module. 
All the programs were compiled with LLVM using O3 compile-time and link-time optimizations.

\subsubsection{General Purpose Registers Reservations}
To implement \sysnamecfi, a minimum of two free GPRs per function is required. 
In the event that an insufficient number of free GPRs is available, additional GPRs must be reserved accordingly. 
To evaluate the feasibility of this requirement, we analyzed all benchmark functions to see how often such register reservations would be necessary. 
Under the default configuration, our results show that 60.55\% of functions have more than two free GPRs, thereby avoiding the need to reserve additional registers.
This indicates that \sysnamecfi's design is practical and applicable across real-world workloads. 
For the remaining functions, standard compiler techniques—such as register spilling to the stack—can be employed with minimal performance impact, ensuring compatibility without compromising correctness or efficiency.

\subsubsection{Runtime Overhead}

Tables~\ref{tab:beebs-rt} and \ref{tab:coremark-pro-rt} present the performance overhead introduced by \sysnamecfi on BEEBS and CoreMark-Pro. 
Overhead is reported as execution time normalized to the baseline.
We compared our work with the state-of-the-art shadow stack solution on ARM Cortex-M, Silhouette shadow stack only (Silhouette SS)~\cite{zhou2020silhouette}.
The performance overhead of \sysnamecfi introduces an average of 7.33\% and 1.81\% on BEEBS and CorMark-Pro, respectively.

For BEEBS benchmark suites, Silhouette SS achieves low overhead in most benchmarks, with an average of 6.24\% and a geometric mean of 0.28\%. 
However, Silhouette SS suffers from high variability. 
It incurs significant slowdowns in specific programs, most notably \texttt{ctl-string} with a 104.34\% overhead and \texttt{qrduino} with -0.01\%, indicating potential performance bottlenecks under certain control-flow patterns.
In contrast, \sysnamecfi maintains more consistent performance across all benchmarks. 
It achieves an average overhead of 7.33\% and a geometric mean of 0.90\%.
While this is slightly higher than Silhouette, in terms of average overhead, 
\sysnamecfi avoids extreme cases of degradation. 
The highest overhead observed is 56.18\% for \texttt{sglib-rbtree}, which is considerably lower than Silhouette SS's worst case. 
Moreover, \sysnamecfi slightly improves performance in some benchmarks, such as \texttt{fir}, which shows a negative overhead of -1.25\%.
WATSON incurs higher overhead than Silhouette SS in some cases due to our compressed shadow stack design, which trades performance for reduced memory footprint. 
This trade-off is intentional for memory-constrained platforms. 
Overall, WATSON achieves competitive and stable performance across benchmarks.

\begin{table}[!t]
	\caption{Performance overhead on BEEBS programs, along with the times (in ms) of baseline, shadow stack (ss) enabled, and silhouette ss.}
	\footnotesize
	\renewcommand{\arraystretch}{1.1}
	\setlength\tabcolsep{0.2ex}
	\label{tab:beebs-rt}
	\begin{tabular}{l|r|rr|rrrrrrr}
		& \multicolumn{1}{c|}{Baseline} & \multicolumn{2}{c|}{Silhouette SS} & \multicolumn{2}{c}{Watson}   \\ \cline{2-8}
		& \multicolumn{1}{c|}{Time} 			& \multicolumn{1}{c}{Time}		& \multicolumn{1}{c|}{Overhead (\%)} 		& \multicolumn{1}{c}{Time}			& \multicolumn{1}{c}{Overhead (\%)}    		\\ \hline \hline
		bubblesort           & 2,755               & 2,757       & 0.07        & 2,764      & 0.33        \\
		ctl-string           & 692                & 1,414       & \textbf{104.34}      & 803       & 16.04        \\
		cubic                & 28,657              & 28,706      & 0.17        & 28,760     & 0.36        \\
		dijkstra             & 40,579              & 40,648      & 0.17         & 41,053     & 1.17        \\
		edn                  & 2,677               & 2,678       & 0.04        & 2,681      & 0.15        \\
		fasta                & 16,274              & 16,274      & 0            & 16,278     & 0.03        \\
		fir                  & 16,418              & 16,419      & 0.01        & 16,212     & \textbf{-1.25}       \\
		frac                 & 8,842               & 8,849       & 0.08        & 8,905      & 0.71        \\
		huffbench            & 46,129              & 46,130      & 0.01        & 46,134     & 0.01        \\
		levenshtein          & 7,806               & 7,873       & 0.86        & 7,965      & 2.04        \\
		matmult-int          & 5,901               & 5,903       & 0.03        & 5,911      & 0.17        \\
		ndes                 & 1,941               & 1,957       & 0.82        & 2,086      & 7.47         \\
		nettle-aes           & 7,024               & 7,029       & 0.07        & 7,042      & 0.26        \\
		picojpeg             & 42,974              & 44,586      & 3.75        & 56,714     & 31.97       \\
		qrduino              & 43,579              & 43,575      & \textbf{-0.01}       & 43,638     & 0.14        \\
		sglib-dllist         & 1,326               & 1,328       & 0.15        & 1,332      & 0.45        \\
		sglib-listinsertsort & 1,359               & 1,360       & 0.07        & 1,364      & 0.37        \\
		sglib-listsort       & 1,058               & 1,059       & 0.10        & 1,064      & 0.57        \\
		sglib-queue          & 2,135               & 2,135       & 0            & 2,140      & 0.23        \\
		sglib-rbtree         & 7,802               & 8,516       & 9.15        & 12,185     & \textbf{56.18}       \\
		stb\_perlin          & 3,164               & 3,399       & 7.43        & 3,792      & 19.85       \\
		trio-sscanf          & 1,264               & 1,389       & 9.89        & 1,567      & 23.97       \\ \hline
		Min                  & 692                & 1,059       & -0.01       & 803       & -1.25       \\
		Max                  & 46,129              & 46,130      & 104.34      & 56,714     & 56.18       \\
		Geomean              & 5,916.25            & 6,200.17    & 0.28       & 6,304.04   & 0.90        \\
		Average              & 13,198              & 13,362.91   & 6.24        & 14,108.64           & 7.33 \\ \hline
	\end{tabular}
\end{table}

For CoreMark-Pro benchmark suites, Silhouette SS demonstrates minimal overhead, with an average of 0.31\% and a geometric mean of 0.15\%. 
For most benchmarks, such as \texttt{linear\_alg}, \texttt{loops-all-mem}, and \texttt{radix2}, Silhouette SS introduces no overhead at all.
The highest observed overhead is only 1.19\% for the core benchmark, confirming that Silhouette SS maintains a low performance impact on complex workloads.
\sysnamecfi exhibits slightly higher overhead, averaging 1.81\% with a geometric mean of 0.59\%. 
While most workloads still demonstrate minimal performance impact, such as \texttt{nnet\_test} with a slight improvement (-0.03\%) and \texttt{linear\_alg} with only 0.01\% overhead.
\sysnamecfi incurs higher costs in certain cases.  
Notably, the core benchmark experiences a 9.44\% overhead, which contributes to the higher overall average. 

\begin{table}[!t]
	\centering
	\caption{Performance overhead on CoreMark-Pro along with the times of baseline, silhouette ss, and \sysnamecfi.}
	\footnotesize
	\renewcommand{\arraystretch}{1.1}
	\setlength\tabcolsep{0.3ex}
	\label{tab:coremark-pro-rt}
	\begin{tabular}{l||r|rr|rrrrrrr}
		& \multicolumn{1}{c|}{Baseline} & \multicolumn{2}{c|}{Silhouette SS} & \multicolumn{2}{c}{Watson}   \\ \cline{2-8}
		& \multicolumn{1}{c|}{Time} 			& \multicolumn{1}{c}{Time}		& \multicolumn{1}{c|}{Overhead (\%)} 		& \multicolumn{1}{c}{Time}			& \multicolumn{1}{c}{Overhead (\%)}    		\\ \hline \hline
		core                       & 136,807    & 138,438    & \textbf{1.19}    & 149,720    & \textbf{9.44}    \\
		linear\_alg-... & 18,277     & 18,277     & \textbf{0}        & 18,278     & 0.01    \\
		loops-all-m...       & 35,239     & 35,239     & \textbf{0}        & 35,239     & 0        \\
		nnet\_test                 & 222,071    & 222,076    & 0.01    & 221,999    & \textbf{-0.03}   \\
		parser-125k                & 9,984      & 10,021     & 0.37    & 10,133     & 1.49    \\
		radix2-...             & 17,269     & 17,269     & \textbf{0}        & 17,269     & 0        \\
		sha-test                   & 40,529     & 40,764     & 0.58     & 41,238     & 1.75    \\
		\hline
		Min                        & 9,984      & 10,021     & 0        & 10,133     & -0.03   \\
		Max                        & 222,071    & 222,076    & 1.19    & 221,999    & 9.44    \\
		Geomean                    & 38,979.66 & 39,098.76 & 0.15   & 39,665.44 & 0.59   \\
		Average                    & 68,596.57 & 68,869.14 & 0.31    & 70,553.71 & 1.81  \\ \hline
	\end{tabular}
\end{table}

\subsubsection{Code Size Overhead}

\begin{figure}[!t]
	\begin{centering}
		\centering
		\includegraphics[width=0.47\textwidth]{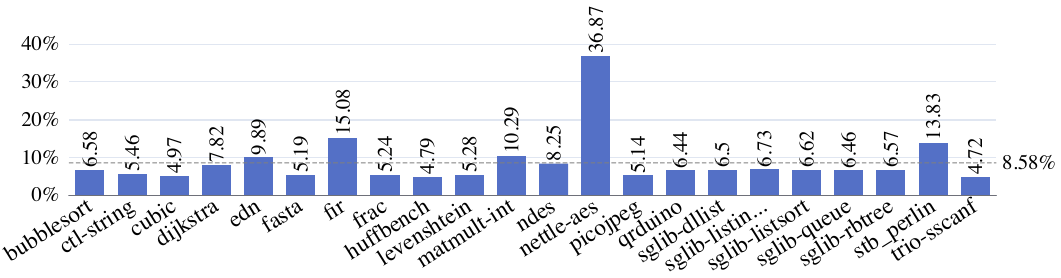}
		\cprotect\caption{Code size overhead of \sysnamecfi on BEEBS.}
		\label{fig:beebs_codesize}
	\end{centering}
\end{figure}

Figure~\ref{fig:beebs_codesize} and Table~\ref{tab:codesize-coremark-pro} present the code size overhead introduced by \sysnamecfi.
We did not compare against Silhouette SS because it uses a parallel shadow stack, whereas \sysnamecfi uses a compact shadow stack, resulting in code size overheads that are not directly comparable.

On BEEBS, \sysnamecfi incurs a geometric mean overhead of 7.36\%, with an average increase of 8.58\% in code size compared to the baseline. 
As described in~\S\ref{subsubsec:microbenchmark}, the \texttt{ssp} register is memory-mapped, and each update operation involves two memory references, leading to instruction expansion.
Thus, a significant portion of this overhead stems from instructions that update the shadow stack pointer. 
Among the BEEBS benchmarks, \texttt{nettle-aes} experiences the highest overhead at 36.87\%, primarily due to the AES round encryption routine.
This benchmark defines each subprocess as an individual function that executes instrumented code, thereby amplifying overhead.

A similar pattern is observed in Table~\ref{tab:codesize-coremark-pro} for CoreMark-Pro, where \sysnamecfi exhibits moderate but acceptable code size overhead, ranging from 2.54\% to 3.53\%.
In contrast, Silhouette SS causes fewer code size overhead between 0.87\% and 1.01\%. 
This level of overhead is expected.
\sysnamecfi incorporates additional logic to maintain a compact shadow stack along with fine-grained memory protection mechanisms. 
However, Silhouette SS maintains a parallel shadow stack without requiring an \texttt{ssp}.

\begin{table}[!t]
	\centering
	\caption{Code size overhead of \sysnamecfi on CoreMark-Pro}
	\footnotesize
	\renewcommand{\arraystretch}{1.1}
	\setlength\tabcolsep{2ex}
	\label{tab:codesize-coremark-pro}
	\begin{tabular}{l|r|rr|rrrrrrr}
		& \multicolumn{1}{c|}{Baseline} & \multicolumn{2}{c|}{Silhouette SS} & \multicolumn{2}{c}{Watson}   \\ \cline{2-8}
		& \multicolumn{1}{c|}{Bytes} 			& \multicolumn{1}{c}{Bytes}		& \multicolumn{1}{c|}{Overhead} 		& \multicolumn{1}{c}{Bytes}			& \multicolumn{1}{c}{Overhead}    		\\ \hline \hline
		Min                  &  52,336             & 52,688      & 0.87        & 53,564     & 2.54        \\
		Max                  &  99,860             & 100,864     & 1.01        & 103,388    & 3.53         \\ \hline	
	\end{tabular}
\end{table}

\subsubsection{Microbenchmark Analysis}
\label{subsubsec:microbenchmark}

To better understand the performance overhead introduced by \sysnamecfi, 
we break down CPU cycles into four types of operations: altering the shadow stack's write protection (AW), updating the shadow stack (USS), adjusting the shadow stack pointer (ASSP), and others. 
We measured the average number of CPU cycles consumed by each type of instruction in our instrumentation. 
Our findings align closely with the ARM reference manual~\cite{ArmCortexM4_ins}, which provides corresponding microarchitectural timing estimates.

\begin{figure}[!t]
	\begin{centering}
		\centering
		\includegraphics[width=0.49\textwidth]{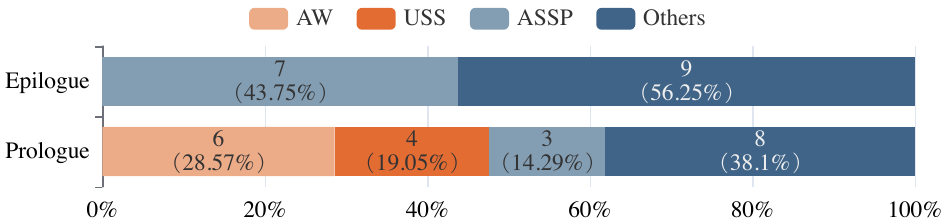}
		\cprotect\caption{CPU cycles breakdown.}
		\label{fig:microbench}
	\end{centering}
\end{figure}

Figure~\ref{fig:microbench} shows that the function prologue phase incurs approximately 21 CPU cycles, with AW, USS, and ASSP accounting for 28.57\%, 19.05\%, and 14.29\% of this overhead, respectively. 
The function epilogue takes 16 CPU cycles, with ASSP being the dominant contributor at 43.75\%. 
This is expected, as the epilogue phase only requires read access to the shadow stack and does not involve changing \sysnamecfi configurations.
These results highlight that the primary sources of overhead stem from fine-grained memory access control and shadow stack pointer adjustments. 

It is also worth noting that \sysnamecfi adopts a compact shadow stack design, in contrast to the parallel shadow stack approach used by Silhouette. 
While our design reduces memory footprint, which is an important advantage on embedded systems, it introduces additional complexity in shadow stack management. 
This represents a deliberate trade-off tailored for resource-constrained environments.

\subsubsection{Exception Handling Overhead}

Table~\ref{tab:interrupt} summarizes the runtime and code size overhead associated with exception handling, specifically focusing on the \textit{Usage Fault} exception. 
We \textbf{exclude} Silhouette from this comparison because it does not support interrupt handling. 
Our evaluation considers three configurations: the baseline without the shadow stack, \sysnamecfi with shadow stack support (\sysnamecfi SS), and \sysnamecfi with both shadow stack and interrupt support enabled (\sysnamecfi Interrupts). 
Due to the small size of the Usage Fault handler, each configuration is executed 100,000 times to amplify measurable differences and demonstrate cumulative overhead across repeated exception handling.

Compared to the baseline, enabling \sysnamecfi introduces a runtime overhead of approximately 9e-06\%, primarily resulting from the additional logic required to maintain the shadow stack during exception entry and return. 
Interestingly, when interrupt support is also enabled, the measured overhead decreases slightly by 2e-05\%. 
This reduction is not due to a performance gain from the added instrumentation, but rather a result of improved code layout and instruction cache alignment, which can occur when additional instructions shift hot paths into more favorable memory alignment. 
In other words, the added interrupt-handling logic incidentally leads to more stable or predictable execution patterns, slightly improving runtime consistency at scale.

In terms of code size, \sysnamecfi increases the binary footprint by 1.79\%, and the addition of interrupt support raises it further to 2.38\%.
This increase reflects the extra instructions inserted into exception handler function entries and exists to preserve and restore the ESF.

\begin{table}[!t]
	\centering
	\caption{Performance and code size overhead associated with exception handling (Usage Fault exception).}
	\footnotesize
	\renewcommand{\arraystretch}{1.1}
	\setlength\tabcolsep{1.8ex}
	\label{tab:interrupt}
	\begin{tabular}{l|rr|rr|rr|rrrrr}
			& \multicolumn{2}{c|}{Baseline} & \multicolumn{2}{c|}{Watson SS}   & \multicolumn{2}{c}{Watson Interrupts}   \\ \hline \hline
			Runtime 		&  	200,047		&& 	200,049		&& 	200,042					 							    \\
			 \hline
			Code Size 		&  	31,506 		&& 	32,070	(1.79\%)	&& 		32,256 (2.38\%)											    \\  \hline 
		\end{tabular}
\end{table}

\subsubsection{Real-world Application Evaluation}
We further evaluate \sysnamecfi using two real-world embedded applications: Pinlock and FatFS\_RAM. 
As Table~\ref{tab:realworld} shows, \sysnamecfi introduces only a 0.2\% performance overhead for Pinlock compared to the baseline, and no measurable overhead for FatFS\_RAM. 
In terms of code size, \sysnamecfi incurs an overhead of 0.88\% for Pinlock and 2.11\% for FatFS\_RAM.
These minimal overheads are largely attributed to the fact that, unlike benchmark suites, such as BEEBS or CoreMark-Pro, 
these applications incorporate HAL libraries and interact with peripherals such as GPIO and memory-mapped storage. 
Since the HAL is assumed to be trusted and is excluded from instrumentation, \sysnamecfi is applied only to the application logic. 
This selective instrumentation significantly reduces the instrumentation footprint, 
demonstrating the practicality and efficiency of \sysnamecfi in realistic embedded system scenarios.

\begin{table}[!t]
	\centering
	\caption{Performance overhead on real world programs.}
	\footnotesize
	\renewcommand{\arraystretch}{1.1}
	\setlength\tabcolsep{0.7ex}
	\label{tab:realworld}
	\begin{tabular}{l|r|r|rr|rrrrrr}
		& \multicolumn{2}{c|}{Baseline} & \multicolumn{4}{c}{Watson}   \\ \cline{2-8}
		& Time 	& Code Size 			& Time		& Overhead    		& Code Size			& Overhead    		\\ \hline \hline
		Pinlock           & 523               & 59,876 & 524      & 0.2       & 60,408 & 0.88\\
		FatFS\_RAM           & 20,480            & 46,260 & 20,480       & 0    & 47,236 & 2.11    \\ \hline
	\end{tabular}
\end{table}

\subsection{Security Evaluation}

To demonstrate the effectiveness of \sysnamecfi in practice, we implemented two attack scenarios where an adversary exploits memory corruption vulnerabilities to hijack control flow. 
Since the real attack involves multiple stages, we focus on the stage of enforcing shadow stack integrity.
Listing~\ref{code:attackscenarios} shows a piece of vulnerable code that merges two attack scenarios.
The vulnerable function \texttt{bar()} copies data from \texttt{user\_input}  (Lines 14) into a fixed-size buffer (Lines 5), allowing the attacker to overwrite local variables and the return address on the main stack. 
In the first attack scenario, the attacker overwrites a function's return address on the main stack, redirecting execution to a crafted payload. 
In the second attack scenario, the attacker leverages knowledge of the system's memory layout to directly tamper with the shadow stack, attempting to bypass control-flow integrity enforcement.
We will explain how \sysnamecfi effectively mitigates both attacks.

\begin{listing}[!h]
	{\renewcommand\fcolorbox[4][]{\textcolor{gray}{\strut#4}}
		\begin{minted}[xleftmargin=5.8pt,numbersep=1pt, tabsize=2, frame=single, framesep=1mm, breaklines, highlightlines={3-4, 6-8, 12, 16-17}, linenos=true, escapeinside=||, fontsize=\footnotesize]{tasm}
	void bar(char *input, uintptr_t *ptr1) {
		char buffer[6];
		|\xglobal\colorlet{FancyVerbHighlightColor}{orange!20}|uintptr_t bar_ptr1;        // for scenario 2
		|\xglobal\colorlet{FancyVerbHighlightColor}{red!20}|strcpy(buffer, input);     // buffer overflow
		
		|\xglobal\colorlet{FancyVerbHighlightColor}{orange!20}|bar_ptr1 = (uintptr_t)ptr1;// for scenario 2
		// Write to where ptr1 points
		*bar_ptr1 = buffer[0];     // attack scenario 2
	}
	void foo() {
		// attacker-controlled input: user_input
		|\xglobal\colorlet{FancyVerbHighlightColor}{red!20}|bar(user_input, (int *)0x20001000);
		printf("Control-Flow Safe!\n");
	}
	void baz() {
		|\xglobal\colorlet{FancyVerbHighlightColor}{blue!20}|// malicious target
		printf("Control-Flow Violation!\n");
	}
	int main() {
		foo();
		return 0;
	}
		\end{minted}
	} 
	\cprotect\caption{Evaluation code for attack scenarios. The goal is to execute \texttt{baz}. Vulnerable code is highlight in red. Code for attack scenario 2 only is highlighted in orange.}
	\label{code:attackscenarios}
\end{listing}
 
In attack scenario 1, the return address is replaced with the address of a malicious function \texttt{baz()}, which is never invoked during benign execution. 
However, during execution, \sysnamecfi restores the correct return address from the shadow stack, ensuring that control returns safely to \texttt{foo()} and preventing the attack from succeeding.
In attack scenario 2, we assume that the attacker has already discovered the address of the shadow stack. 
The attacker attempts to overwrite a shadow stack entry directly (Listing~\ref{code:attackscenarios}, Line 9) to redirect execution to \texttt{baz()} upon returning from \texttt{bar()}. 
However, the write protection of \sysnamecfi blocks this unauthorized access and triggers an exception.
It is important to note that the second attack scenario is designed to evaluate the effectiveness of \sysnamecfi's write protection for the shadow stack. 
While the attack does overwrite local variables, its primary goal is not to demonstrate a defense against data-only attacks, but rather to test whether unauthorized writes to the shadow stack are reliably detected and blocked. 

Our evaluation confirms that any attempt to write to the shadow stack outside the permitted prologue and epilogue phases reliably triggers an exception, effectively preserving the integrity of return addresses and preventing tampering.

\begin{table*}[!t]
	\centering
	\footnotesize
	\caption{Comparison with the state-of-the-art approaches.}
	\renewcommand{\arraystretch}{1.1}
	\setlength\tabcolsep{0.08ex}	
	\begin{threeparttable}
		\begin{tabular}{l|cc|ccc|ccc|ccccccccc} \hline
			& \multicolumn{2}{c|}{Enforcement Policy} & \multicolumn{3}{c|}{Prototype} & \multicolumn{3}{c|}{Protect Targets} & \multicolumn{4}{c}{Compatibility}\\ \cline{2-15}
			& \begin{tabular}[c]{@{}c@{}}Backward\end{tabular} & \begin{tabular}[c]{@{}c@{}}Required \\Hardware\end{tabular} & Arch. & System & Overhead & \begin{tabular}[c]{@{}c@{}}Unpriv.\\Code\end{tabular} & \begin{tabular}[c]{@{}c@{}}Exception \\/Interruption \end{tabular}  &\begin{tabular}[c]{@{}c@{}} Other \\ Priv. Code\end{tabular} & \begin{tabular}[c]{@{}c@{}}Privilege\\Separation\end{tabular} & \begin{tabular}[c]{@{}c@{}}Compartmen-\\talization\end{tabular} & TEE & \begin{tabular}[c]{@{}c@{}}Support\\Arch.\end{tabular}\\ 
			\hline\hline
			RECFISH~\cite{walls2019control} 			& shadow stack 							& MPU 					& R & RTOS & 21\% 						& $\testmark$	& $\times$ 				& $\times$ 		& $\times$ & $\times$ & $\testmark$ & Universal\\ 
			$\mu$RAI~\cite{almakhdhubmurai2020}		& return address integrity 	& MPU						& M & Bare-metal & 0.1\% 			& $\testmark$	& $\testmark$ 		& $\testmark$	& $\times$ & $\times$ & $\testmark$ & Universal\\
			SUM~\cite{choi2024sum}								& shadow stack 							& MPU   				& M & Both & 2.77\% 					& $\testmark$	& $\testmark$ 		& $\testmark$	& $\times$ & $\times$ & $\testmark$ & Universal\\
			Silhouette~\cite{zhou2020silhouette}	& shadow stack 							& ULSI/MPU 			& M & Bare-metal & 3.4\% 			& $\times$		&$\times$					& $\testmark$	& $\times$ & $\times$ & $\testmark$ & Universal\\
			Kage~\cite{du2022kage} 								& shadow stack 							& ULSI/MPU 			& M & RTOS& 5.2\% 						& $\testmark$	& $\halfcheckmark$& $\testmark$	& $\times$ & $\times$ & $\testmark$ & Universal\\
			\hline
			CaRE~\cite{nyman2017cfi} 							& shadow stack 							& TrustZone  					& M & Bare-metal& 513\% 			& $\testmark$	& $\testmark$			& $\testmark$	& $\testmark$ & $\testmark$ & $\halfcheckmark$ & Selected\\
			TzmCFI~\cite{kawada2020tzmcfi} 				& shadow stack 							& TrustZone  					& M & RTOS& 84\% 							& $\testmark$	& $\testmark$			& $\testmark$	& $\testmark$ & $\testmark$ & $\halfcheckmark$ & Selected\\
			RIO~\cite{kim2023rio}									& encrypt return addresses  & TrustZone   					& M & Bare-metal & 16.83\%		& $\testmark$	& $\testmark$ 		& $\testmark$	& $\testmark$ & $\testmark$ & $\halfcheckmark$ & Selected\\
			\hline
			Sherloc~\cite{tan2023sherloc}					& shadow stack 							& TrustZone/MTB 				& M & Both &$>$100\% 								& $\testmark$	& $\testmark$ 		& $\testmark$	& $\testmark$ & $\testmark$ & $\halfcheckmark$ & Restricted\\
			
			InsectACIDE~\cite{wang2024insectacide}					& shadow stack 							& TrustZone/MTB 				& M & RTOS &- 								& $\testmark$	& $\testmark$ 		& $\testmark$	& $\testmark$ & $\testmark$ & $\halfcheckmark$ & Restricted\\
			DeTRAP~\cite{richter2024detrap}				& shadow stack 							& Debug Trigger & RISC-V 	 &  Bare-metal & 1.9\%	&	$\testmark$ &	$\testmark$ 		& $\testmark$ & $\testmark$ & $\testmark$ & $\testmark$ & Universal\\
			\hline
			\textbf{\sysnamecfi} 									& shadow stack 							& DWT 					& M & Bare-metal & 7.33\% 		& $\testmark$	& $\testmark$ 		& $\testmark$	& $\testmark$ & $\testmark$ & $\testmark$ & Universal\\
			\hline                                                                
		\end{tabular}
		\begin{tablenotes}
			\item[] 
			Legend:
			ULSI: unprivileged load or store instructions. 
			Selected/restricted compatibility: the approach requires one/more special extensions that are not widely supported.
			M: Cortex-M, R: Cortex-R.
		\end{tablenotes}
	\end{threeparttable}	
	\label{tab:cfi-comparison}
\end{table*}

\subsection{Comparison with DeTRAP}

DeTRAP~\cite{richter2024detrap} is the most closely related work that utilizes hardware debug features for shadow stack protection. 
However, it is tailored for RISC-V, overlooking issues faced by other architectures. 
First, DeTRAP reserves GPRs to accelerate shadow stack access.
This design choice is viable on RISC-V, which provides 32 GPRs
However, for other microcontroller architectures with even fewer registers, such as Cortex-M, which only has 13 (R0-R12), reserving GPRs can lead to significant performance degradation.
Moreover, when uninstrumented functions inadvertently use these reserved registers, resolving register conflicts and ensuring correct shadow stack enforcement becomes significantly more complex.
In contrast, \sysnamecfi reserves the shadow stack pointer in memory-mapped DWT registers, avoiding GPR reservation and its associated issues.

Second, DeTRAP relies on RISC-V's exception model with a centralized trusted handler that can intercept all exceptions via a single entry point.
However, Cortex-M adopts a fundamentally different exception model that prioritizes efficiency via hardware-assisted stacking and unstacking. 
This model lacks a centralized trusted exception handler, rendering DeTRAP's approach inapplicable.
\sysnamecfi on Cortex-M modifies the exception handler's prologue and epilogue instead of introducing an additional trap handler.

Overall, our contribution should not be viewed as an implementation variant of DeTRAP. 
The architectural differences fundamentally alter the available enforcement mechanisms and invalidate key assumptions underlying prior designs. 
For the same reason, we argue that direct performance comparisons between DeTRAP and \sysnamecfi are unfair, as the two systems operate under substantially different architectural constraints.

\section{Security Analysis}
\label{sec:sec_analysis}

This section analyzes how \sysnamecfi mitigates backward-edge control-flow hijacking attacks under the threat model defined before. 

\emph{Security of Backward-edge Control-flow Protection.} 
To hijack control flow, an attacker may attempt to overwrite the return address on the stack. 
\sysnamecfi defends against this by maintaining a compact shadow stack that stores a protected copy of each return address. 
\sysnamecfi also protects the backward edges introduced by exceptions/interrupts.
As a result, even if the main stack is compromised, the control flow cannot be redirected.

\emph{Security of Exception Control-flow Protection.}
Exception handlers, such as those triggered by Usage Faults, implicitly introduce additional backward edges by saving execution contexts. 
If left unprotected, these saved exception frames can be exploited to hijack control flow. 
To address this, \sysnamecfi extends shadow stack protections to cover exception entry and return. 
This closes a critical gap in conventional shadow stack designs by preventing forged exception return addresses.

\emph{Security of Shadow Stack Protection.}
An attacker may attempt to tamper with the shadow stack directly. 
To prevent this, \sysnamecfi enforces fine-grained memory protection: the shadow stack region is marked read-only by default, and write access is granted only during legitimate updates. 
This ensures that even attackers with arbitrary memory write capabilities cannot modify return addresses. 
Moreover, \sysnamecfi incorporates LLVM-based forward-edge CFI to block indirect transfers and prevent reuse of routines to disable the write protection of the shadow stack.

\emph{Security of Shadow Stack Pointer Protection.}
The shadow stack pointer holds the address of the current top of the shadow stack and is essential for validating return addresses. 
If an attacker were able to corrupt this pointer, they could redirect it to a memory region under their control and construct a fake shadow stack to bypass the shadow stack protection.
An attacker may attempt to corrupt the pointer in one of two ways. 
First, by overwriting a return address (backward-edge attack) and attempting to redirect control flow to a routine that updates the pointer. 
However, \sysnamecfi enforces return address integrity to prevent such tampering. 
Second, the attacker may try to mount a forward-edge attack by corrupting a function pointer or control transfer target to reach code that legitimately updates the shadow stack pointer. 
To prevent this, we apply LLVM's forward-edge CFI, which restricts indirect branches to valid, statically verified targets. 

\section{Related Work}
\label{sec:related}

There has been a considerable number of mechanisms designed to protect the embedded systems.
Since we focus on the return address integrity, we summarize the current state-of-the-art in Table~\ref{tab:cfi-comparison} and discuss them in the following.

Among those mechanisms, the memory management technique (e.g., MPU) has been widely used.
For instance, RECFISH~\cite{walls2019control} utilizes the MPU to configure the shadow stack in privileged memory to protect unprivileged tasks. 
It modifies the scheduler to store task states in shadow stacks to support multiple tasks.
Silhouette~\cite{zhou2020silhouette} and Kage~\cite{du2022kage} are the most efficient shadow stack implementations for bare-metal systems and RTOS, respectively. 
They also use the MPU to enforce privileged-only access to the shadow stack, 
transforming regular store instructions of the untrusted code into unprivileged ones to prevent unauthorized writes to the shadow stack.
Kage further modifies the scheduler to manage shadow stacks for tasks and uses a trusted dispatcher for exception handling. 
SUM~\cite{choi2024sum} protects the shadow stack by restricting all access through the MPU and temporarily disabling the MPU via FaultMask.

However, those approaches above are incompatible with existing protection mechanisms that also utilize the MPU, as they face limitations of configurable MPU regions~\cite{tan2024sok}. 
\sysnamecfi does not depend on memory management techniques but repurposes the debugging units to protect the shadow stack.
Moreover, \sysnamecfi is compatible with the original use of the debugging units.

Other than a shadow stack, there are other solutions to ensure the return address integrity. 
For instance, $\mu$RAI~\cite{almakhdhubmurai2020} removes the need to spill the return address to the stack.
It uses direct jump instructions, a reserved general-purpose register, and statically computed return address lookup tables to determine the correct return location at run-time. 
It handles exceptions by modifying the interrupt service routine to save registers to a safe region.
However, it requires a complete static call graph, whereas \sysnamecfi avoids the complexity with lower code memory overhead.

Some mechanisms leverage rarely implemented security extensions to protect the shadow stack.
Both CaRE~\cite{nyman2017cfi} and TzmCFI~\cite{kawada2020tzmcfi} use TrustZone for this purpose, targeting bare-metal systems and RTOS, respectively.
They put the shadow stack in the TrustZone secure world, but incur high performance overhead, which is up to 513\%.
SHERLOC~\cite{tan2023sherloc} and InsectACIDE~\cite{wang2024insectacide} are control-flow violation detection approaches that use a hardware trace unit to reconstruct and validate control flows in the TrustZone secure state. 
\sysnamecfi introduces less performance overhead and does not require rarely featured hardware units.

There are other related projects that repurposed hardware debug units. 
For example, PicoXoM~\cite{shen2020fast} configures the watchpoints to implement execute-only memory (XOM).
Jang et al.~\cite{jang2019process} use watchpoints to isolate the user memory from the kernel memory and set the kernel memory as execute-only.
DeTRAP configures the debug trigger on RISC-V to monitor write access to the shadow stack and implements trap handling to interpret all exceptions. 

\section{Limitations and Discussion}

In this section, we discuss the limitations of our current design and implementation, as well as future improvements.

\emph{Challenges in Protecting Third-party Libraries.}
In theory, \sysnamecfi can instrument third-party libraries for more comprehensive protection, as they may contain vulnerabilities. 
However, protecting them is difficult due to their pre-compiled nature, lack of source visibility, relocation complexity, and potential ABI violations. 
Recompilation is labor-intensive and requires intrusive modifications. 
Alternative strategies, such as whole-program compilation, IR instrumentation, dynamic binary rewriting, or wrapper protection, have trade-offs. 
Future work can apply heuristic analysis to source code and selectively modify vulnerable functions to add shadow stack protection.

\emph{Optimized Shadow Stack.}
While \sysnamecfi instruments all functions with return instructions for complete backward-edge protection, not all require such strict enforcement. 
Functions that do not allocate local stack storage or receive external input are less vulnerable to return address corruption. 
Inspired by \texttt{\_FORTIFY\_SOURCE} in the glibc library~\cite{gcc_fortify_src}, we propose a heuristic-driven optimization for future work. 
This would selectively instrument potentially vulnerable functions based on static analysis.
\sysnamecfi could include configurable levels 
A \textit{lightweight mode} protects only high-risk functions, while a \textit{full mode} maintains comprehensive coverage. 
This strategy would further reduce performance and size overhead, making the approach more practical for resource-constrained embedded systems.

\emph{Size of the Shadow Stack on ARMv7-M.}
Due to the limited number of configurable DWT comparators on ARMv7-M, only 32 KB of memory can be protected using a single comparator. 
Our shadow stack design occupies this 32 KB region, which—assuming 4 bytes per return address—can support up to 8192 protected returns. 
We consider this size sufficient for two reasons: 
(1) although modular design is encouraged, excessive use of small, highly fragmented functions that introduce frequent branching is discouraged in embedded systems due to performance constraints, 
and (2) embedded applications typically do not involve deep recursive call stacks with thousands of nested function calls.

\emph{Compatibility with Debugging Purpose and Security Mechanisms.} 
The DWT unit is commonly used for profiling and debugging, with \texttt{COMP0} frequently reserved for CPU cycle counting. 
As shown in our evaluation (\S\ref{sec:eva}), \sysnamecfi relies on \texttt{COMP0} for cycle counting while preserving the shadow stack's write protection. 
This demonstrates that \sysnamecfi remains compatible with standard DWT functionality and does not interfere with typical debugging workflows.
Other security mechanisms also make use of the DWT. 
For instance, picoXoM~\cite{shen2020fast} leverages DWT comparators to implement XOM. 
However, newer Cortex-M processors now provide built-in XOM support, eliminating the need to repurpose DWT resources for this protection.

\section{Conclusion}

This paper presents \sysnamecfi, a lightweight and practical shadow stack enforcement mechanism designed for embedded systems. 
Unlike prior solutions, \sysnamecfi does not depend on a memory management unit or rare security extensions.
\sysnamecfi offers backward-edge protection through a compact shadow stack with fine-grained write protection and extends this to exception handling.
It further strengthens security by integrating LLVM-based forward-edge CFI.
\sysnamecfi incurs an average runtime overhead of 7.33\% on the BEEBS benchmark and 1.81\% on CoreMark-Pro, with negligible overhead in exception handling and two real-world applications.
Overall, \sysnamecfi is \textit{effective} and \textit{efficient} across various embedded systems.


\bibliographystyle{IEEEtran}
\bibliography{ref}

\newpage


\end{document}